\documentclass{article}
\usepackage{graphicx}  

\newcommand{\numu}{\mbox{$\nu_{\mu}$}}
\newcommand{\nutau}{\mbox{$\nu_{\tau}$}}
\newcommand{\nualpha}{\mbox{$\nu_{\alpha}$}}
\newcommand{\nui}{\mbox{$\nu_{\mathrm i}$}}

\newcommand{\mnue}{\mbox{$m(\nu_{\mathrm e})$}}
\newcommand{\mtwonue}{\mbox{$m^2(\nu_{\mathrm e} )$}}
\newcommand{\mnumu}{\mbox{$m(\nu_{\mu} )$}}
\newcommand{\mtwonumu}{\mbox{$m^2(\nu_{\mu} )$}}
\newcommand{\mnutau}{\mbox{$m(\nu_{\tau} )$}}
\newcommand{\mtwonutau}{\mbox{$m^2(\nu_{\tau} )$}}
\newcommand{\mnui}{\mbox{$m(\nu_{\mathrm i})$}}

\newcommand{\mnuone}{\mbox{$m(\nu_{\mathrm 1})$}}
\newcommand{\mnutwo}{\mbox{$m(\nu_{\mathrm 2})$}}

\newcommand{\mnubari}{\mbox{$m(\bar{\nu}_{\mathrm i})$}}
\newcommand{\mnuj}{\mbox{$m(\nu_{\mathrm j})$}}
\newcommand{\mtwonui}{\mbox{$m^2(\nu_{\mathrm i})$}}
\newcommand{\mtwonuj}{\mbox{$m^2(\nu_{\mathrm j})$}}
\newcommand{\mtwonu}{\mbox{$m^2(\nu)$}}
\newcommand{\deltatwomnuij}{\mbox{$\Delta m^2_{\rm ij}$}}
\newcommand{\ualphai}{\mbox{$U_{\mathrm \alpha i}$}}

\newcommand{\rhenium}{\mbox{$^{187}$Re}}

\newcommand{\indexi}{\mbox{$_{\mathrm i}$}}
\newcommand{\indexj}{\mbox{$_{\mathrm j}$}}
\newcommand{\ttwo}{\mbox{$\mathrm T_2$}}
\newcommand{\ev}{\mbox{$\mathrm eV$}}
\newcommand{\evtwo}{\mbox{$\mathrm eV^2$}}
\newcommand{\bdec}{\mbox{$\beta$~decay}}
\newcommand{\bspec}{\mbox{$\beta$~spectrum}}
\newcommand{\belec}{\mbox{$\beta$~electron}}
\newcommand{\bemit}{\mbox{$\beta$~emitter}}
\newcommand{\ezero}{\mbox{$E_0$}}

\newcommand{\mee}{\mbox{$m_{\mathrm ee}$}}
\newcommand{\etal}{\mbox{\it et al.}}

\newcommand{\PRL}{\mbox{Phys. Rev. Lett.}}
\newcommand{\PL}{\mbox{Phys. Lett.}}

\title{Direct neutrino mass search}
\author{Ch. Weinheimer\\
  \normalsize Helmholtz-Institut f\"ur Strahlen- und Kernphysik,\\
  \normalsize Rheinische Friedrich-Wilhelms-Universit\"at,\\ 
  \normalsize D-53115 Bonn, Germany\\
  \normalsize Email: weinheimer@iskp.uni-bonn.de}

\begin{document}

\maketitle

\begin{abstract}
With the compelling evidence for massive 
neutrinos from recent $\nu$-oscillation
experiments, one of the most fundamental tasks of particle physics 
over the next years will be the determination of the absolute 
mass scale of neutrinos,
which has crucial implications for cosmology,
astrophysics and particle physics.
Neutrino oscillation experiments can measure squared mass differences
but not masses. The latter have to be determined 
in a different way.
The direct mass experiments investigate 
-- besides time-of-flight measurements -- the kinematics
of weak decays obtaining information on the neutrino mass without further
requirements. Here the tritium \bdec\ experiments give the most stringent 
results.  
The current tritium \bdec\ experiments at Mainz and Troitsk
are reaching their sensitivity limit. 
The different options for a next generation direct neutrino mass experiment
with sub-eV sensitivity are discussed. The KATRIN experiment,
which will investigate the tritium $\beta$ spectrum with a
MAC-E-Filter of 1~eV resolution, 
is being prepared to reach a sub-eV sensitivity.
\end{abstract}

\section{Introduction}
\label{sec_lablim_intro}  
Recent experimental results from atmospheric and solar neutrinos
give strong evidence that neutrinos oscillate from one flavor
state into another. A neutrino of one
specific flavor eigenstate \nualpha\ is a non-trivial
superposition of different neutrino mass states \nui\
\footnote{Assuming CPT-invariance, here we do not distinguish between masses
of neutrinos \mnui\ and the corresponding antineutrinos \mnubari .}. This fact
is described by
an  unitarity mixing matrix \ualphai
\begin{equation}
  \label{eq_lablim_numixing}
  \nualpha = \sum_i \ualphai \nui \; .
\end{equation}
Future oscillation experiments will determine the elements
\ualphai\ with great precision.

However, $\nu$--oscillations experiments do not yield the values
of the neutrino masses,  they are only sensitive to differences
between squared neutrino masses $\deltatwomnuij =
|\mtwonui-\mtwonuj |$ \footnote{Neutrino oscillation effects involving
matter effects are in principle able to determine the sign of 
\mtwonui-\mtwonu .}. The values $\deltatwomnuij$ from
oscillation experiments only give lower limits on neutrino masses
\begin{equation}
{\rm max}\left(\mnui,\mnuj \right) \geq \sqrt{\deltatwomnuij} \,.
\end{equation}

On the other hand, if the absolute value of one  mass eigenstate
\nui \ is known, all other neutrino masses can be reconstructed
with the help of the differences of the squared neutrino masses,
if the signs of the different $\mtwonui-\mtwonuj$ values are known.

The information on the neutrino mass scale is of crucial importance
for particle physics as well as for astrophysics and cosmology.
The neutrino mass states can be arranged in a
hierarchical way like the charged fermions. This would mean that
the different neutrino masses are
essentially governed by the square root of $\deltatwomnuij$. 
On the other hand the neutrino masses could be quasi-degenerate 
with about the same value -- {\it e.g.} a few tenth of an eV -- and
small mass differences between the different states to explain 
the oscillation signal.
Which neutrino mass scenario
\footnote{Of course any neutrino mass
scenario in between the hierarchical and the
quasi-degenerate pattern would be possible.} 
is right is a very important information
in order to find the right theory beyond the Standard Model.
Big bang theories predict, that  
a huge abundance of relic neutrinos
of all flavors similar to the photons of the cosmic
microwave background radiation (CMBR) exists in the universe. 
An average neutrino mass of 1~eV would contribute to the
energy and matter distribution of the universe
by 8~\% in units of the critical density 
Therefore any neutrino mass of this order would contribute significantly
to the missing dark matter and would influence the evolution of the
universe and its structure formation.

Therefore, 
information on the neutrino masses can also be deduced from
cosmology. Independent information on  neutrino
masses comes from the measurement of the arrival time distribution
of neutrinos emitted in a supernova explosion, the present limit
from SN1987a is $\mnue < 23~\ev$ \cite{pdg00} (see section
\ref{sec_lablim_kin} for the remarks on \mnue ).

Direct information on neutrino masses can be obtained by
laboratory experiments using two different approaches: the
investigation of the decay kinematics of weak decays and the
search for  neutrinoless double \bdec . Both methods  give
complementary information on the neutrino masses \mnui .

This paper is organized as follows: Section 2 describes the direct neutrino
mass measurements, section 3 concentrates on tritium \bdec\ experiments.
In section 4 the options for future neutrino mass
searches is discussed. The planned next generation tritium \bdec\ experiment
KATRIN is presented in section 5. Section 6 gives a short summary.
 
\section{Decay Kinematics of Weak Decays}
\label{sec_lablim_kin} The investigations of the kinematics of
weak decays is based on measurements of the
charged decay products. Using energy and momentum conservation the
missing neutrino mass can be reconstructed from the kinematics of
the charged particles.  The part of the phase space which is most
sensitive to the neutrino mass is the one which corresponds to the
emission of a non-relativistic massive neutrino. Therefore decays
releasing charged particles with  a small free kinetic energy are
preferred.

In principle, a kinematical neutrino mass measurement yields
information on the different mass eigenstates \mnui , since it
performs a projection on energy and mass. But usually the
different neutrino mass eigenstates cannot be resolved by the
experiment. Therefore an average over neutrino mass eigenstates is
obtained which is specific for the  flavor of the weak decay and
hence termed \mnue , \mnumu\ or \mnutau,
respectively\footnote{This average value is not a unique quantity
but depends also on how the experiment is analyzed, in particular
if it is done under the assumption of a single neutrino mass
state. Considering however the small differences of squared masses
\deltatwomnuij\ obtained by neutrino oscillation experiments and comparing
them to the experimental resolutions of the present kinematic
measurements, this question appears to be of a more academic
nature.}. This fact will be discussed in more detail for the case
of the muon neutrino \numu .

\subsection{\mnumu}
The muon neutrino mass \mnumu\ has been investigated in the
two-body decay of a pion at rest:
\begin{equation}
  \label{eq_pidecay}
  \pi^+ \rightarrow \mu^+ + \nu_\mu \quad {\mathrm or} \quad
  \pi^- \rightarrow \mu^- + \bar{\nu}_\mu
\end{equation}
Energy and momentum conservation result in a sharp muon momentum
$p(\mu)$ from which the mass of the muon neutrino \mnumu\ would
follow as:
\begin{equation}
  \label{eq_mnumu}
  \mtwonumu = m^2(\pi) + m^2(\mu)
              - 2 \cdot m(\pi) \cdot \sqrt{m^2(\mu) + p^2(\mu)}
\end{equation}
Eq. (\ref{eq_mnumu}) only holds, if the muon neutrino \numu\ is a
well-defined mass eigenstate, which does not apply in the case of
neutrino mixing. Hence, if the muon momentum $p(\mu)$ for pion
decay at rest could be measured with sufficient precision one
would detect three different values $p^2_{\mathrm i}(\mu)$ with
relative fraction $|U^2_{\mathrm \mu i}|$ corresponding to 
the measurements of the corresponding squared mass values \mtwonui\
contributing to the muon neutrino
\numu . Up to now no  direct neutrino mass measurements has
discriminated  different neutrino masses or has established a
signal for any non-zero neutrino mass.

Therefore on the left part of eq. (\ref{eq_mnumu}) a mean squared
average of mass eigenstates of the muon neutrino is defined as
\begin{equation}
  \label{eq_mnumu_average}
  \mtwonumu = \sum_i | U^2_{\mathrm \mu i} | \cdot \mtwonui
\end{equation}
To deduce \mtwonumu\ from (\ref{eq_mnumu_average}) three
quantities have to be measured with very high precision: 
the muon mass $m(\mu) = 105.6583568(52)$~MeV \cite{pdg00},
the pion mass $m(\pi) = 139.570180 (350)$  bMeV \cite{pdg00},
$m(\pi) = 139.570180 (350)$~MeV, 
and the muon momentum from pion decay
at rest $p(\mu) = 29.791998(110)$~MeV, which has been derived in a
dedicated experiment at Paul Scherrer Institute (Z\"urich)
\cite{pmu}. Putting these values into equation (\ref{eq_mnumu})
one  obtains \cite{pmu}:
\begin{equation}
  \label{eq_mtwonumu_value}
  \mtwonumu = -0.016 \pm 0.023 ~ {\mathrm MeV^2}
\end{equation}
from which an upper limit on the muon neutrino mass can be derived \cite{pdg00}
\begin{equation}
  \label{eq_mnumu_limit}
  \mnumu < 190 ~ {\mathrm keV \quad (90~\%~C.L.)}
\end{equation}

\subsection{\mnutau}
\label{sec_lablim_kin_nutau} The most
sensitive information on \mnutau\ comes from the investigation of
$\tau$ pairs produced at electron-positron colliders decaying into
multi pions. Due to the large mass of the $\tau$ , decays into 5
and 6 pions give the highest sensitivity because they restrict the
available phase space of the \nutau . However,  the corresponding
branching ratios are rather small.

The quantity looked at is the invariant mass of the multiple pions
$M_\pi$. Although $M_\pi$ does not have a direct physical meaning,
the mass of the tau neutrino \mnutau\ restricts $M_\pi$ \ due to
energy and momentum conservation.  In the rest frame of the
decaying $\tau$ \ lepton, $M_\pi^2$ \ is expressed by:
\begin{eqnarray}
  M^2_\pi & = &\Big( \sum_{\mathrm j} E_{\mathrm j}(\pi),
                     \sum_{\mathrm j} \vec{p_{\mathrm j}}(\pi) \Big) ^2
          = \Big( m(\tau)-E(\nutau), -\vec{p}(\nutau) \Big) ^2\\
          & \leq & \Big( m(\tau) - \mnutau \Big)^2
\end{eqnarray}
The most sensitive investigation comes from the ALEPH experiment
at LEP. Its two-dimensional analysis in the
$M_\pi$--$\sum_{\mathrm j} E_{\mathrm j,~lab}(\pi)$ plane
restricts \mnutau \ \cite{aleph}:
\begin{equation}
  \label{eq_mnutau_limit}
  \mnutau < 18.2 ~ {\mathrm MeV \quad (95~\%~C.L.)}
\end{equation}
A further improvement based on  data from B-factories can be
expected with an estimated  sensitivity limit of 3~MeV.

Using the most recent results on atmospheric and solar neutrino
oscillation (see chapter 5) the neutrino mixing matrix \ualphai\
and the squared mass differences \deltatwomnuij\ suggest that the
averages \mtwonumu\ and \mtwonutau\ (compare  eq.
(\ref{eq_mnumu_average})) are rather close due to the strong
\numu\ - \nutau\ mixing and the very small difference $\Delta
m^2_{23}$. Therefore \mnutau\ is already constrained by the limit
on the muon neutrino mass (\ref{eq_mnumu_limit}).

\subsection{\mnue}
\label{sec_lablim_kin_nue}
The mass of the electron neutrino is determined by the investigation
of the electron energy spectrum (\bspec ) of a nuclear \bdec\
\cite{robertson,holzschuh,wilkerson}.
In a $\beta^-$ decay
\begin{equation}
  \rm (Z,A) \rightarrow (Z+1,A)^+ + e^- + \bar{\nu}_e
\end{equation}
the available energy is shared between the $\beta$--electron and
the electron antineutrino, because the recoiling nucleus
practically receives no kinetic energy due to its much heavier
mass. The phase space region of non-relativistic neutrinos, where
the highest  sensitivity to the neutrino mass is achieved,
corresponds to the very upper end of the \bspec . To maximize this
part, a \bemit\ with a very low endpoint energy \ezero\ is
required. This requirement is fulfilled  by \rhenium\ and tritium
(T or $^3$H), which have the two lowest endpoint energies of
$\ezero = 2.6$~keV and $\ezero = 18.6$~keV, respectively.

Although tritium has a higher endpoint energy as compared to
\rhenium\, its use has several advantages:
\begin{itemize}
  \item Tritium decays by a super-allowed transition
    into its mirror nucleus $^3$He resulting in a half life of 12.3 years,
    compared to the primordial half life of the forbidden transition
    of \rhenium\ of $5 \cdot 10^{10}$~a.
    The short half life yields a high specific activity and minimizes
     the inelastic
    processes of \belec s within the tritium source.
  \item Due to the super-allowed decay
    the transition matrix element does not depend
    on the electron energy: the \bspec\ is determined entirely by the
    available phase space.
  \item Tritium has the simplest atomic shell minimizing the necessary
     corrections due to the electronic
     final states or inelastic scattering in the $\beta$ source.
\end{itemize}
These arguments clearly favor tritium for standard setup, which consists
of a $\beta$ source connected to a $\beta$ spectrometer (sometimes
called ``passive source'' setup).
The advantage of the lower \rhenium\ endpoint energy can only be
exploited if
the $\beta$ source and the spectrometer are identical
(sometimes called ``active source'' setup), which is
realized in the case of cryogenic bolometers for instance.

For an allowed or super-allowed transition the electron energy
spectrum is given by Fermi's Golden Rule
\begin{eqnarray}
  \frac{dN}{dE} & = & \frac{G_{\mathrm F}^2}{2 \pi^3 \hbar^7}
                  \cdot \cos^2\Theta_{\mathrm C} \cdot | M |^2 \cdot F(E,Z+1)
                  \cdot p \cdot (E + m) \cdot \varepsilon \nonumber \\
               & ~ & \cdot \sqrt{\varepsilon^2 - \mtwonue}
                     \cdot \Theta(\varepsilon - \mnue) \nonumber \\
                & = & A \cdot F(E,Z+1) \cdot
                  p \cdot (E + m) \cdot \varepsilon \nonumber\\
                & ~ & \cdot
               \sqrt{\varepsilon^2 - \mtwonue}
               \cdot \Theta(\varepsilon - \mnue)
  \label{eq_betaspec}
\end{eqnarray}
with the Fermi coupling constant $G_{\mathrm F}$, 
the Cabbibo angle $\Theta_{\mathrm C}$, the nuclear
transition matrix element $M$ (tritium: $|M|^2 = 5.55 \cdot \hbar^6$
\cite{robertson}), the Fermi function $F$  describing the final
electromagnetic interaction of the emitted \belec\ with the
daughter nucleus $F(E,Z+1)$, the electron mass, momentum and kinetic
energy $m$, $p$ and $E$, and the energy difference $\varepsilon =
\ezero -E$. The Fermi function is approximately given by
\cite{holzschuh}
\begin{equation}
  F(E,Z+1) = \frac{2\pi \eta}{1 - exp( -2\pi \eta )}
\end{equation}
with the Sommerfeld parameter $\eta = \alpha (Z+1) / \beta$.
Equation (\ref{eq_betaspec}) only holds for the decay of a bare
nucleus. For the more realistic case of an atom or  a molecule the
possible excitation of the electron shell due to the sudden change
of the nuclear charge by one unit has to be taken into account.
 The atom or molecule will end in a
specific state of excitation energy V\indexj\ with a
probability W\indexj . The corresponding excitation probabilities
can be calculated in the sudden approximation from the overlap of
the primary electron wave function $\Psi_0$ with the wave
functions of the daughter  ion $\Psi_{\mathrm f,j}$
\begin{equation}
  W\indexj = |\left< \Psi_0| \Psi_{\mathrm f,j} \right>| ^2
\end{equation}
Equation (\ref{eq_betaspec}) is thus modified  into a sum of
$\beta$ spectra of amplitude  W\indexj\ with different endpoint
energies $E_{\mathrm 0,j} = \ezero - V_{\mathrm j}$
\begin{eqnarray}
  \frac{dN}{dE} & = & A \cdot F(E,Z+1) \cdot
                  p \cdot (E + m)  \nonumber\\
                &~& \cdot \sum_{\mathrm j} W\indexj \cdot \varepsilon\indexj
                    \cdot \sqrt{\varepsilon\indexj^2 - \mtwonue}
                    \cdot \Theta(\varepsilon\indexj   - \mnue)
  \label{eq_betaspec_finalstates}
\end{eqnarray}
The energy differences $\varepsilon\indexj$ are then defined as
$\varepsilon\indexj = \ezero - V\indexj - E$.

In case of neutrino mixing the spectrum is a sum of the components
of decays into mass eigenstates
\begin{eqnarray}
  \frac{dN}{dE} & = & A \cdot F(E,Z+1) \cdot
                  p \cdot (E + m)  \nonumber\\
                &~& \cdot \sum_{\mathrm j} W\indexj \cdot \varepsilon\indexj
                    \cdot \left( \sum_{\mathrm i} |U_{\mathrm ei}|^2 \cdot
                                 \sqrt{\varepsilon\indexj^2 - \mtwonui}
                                \cdot \Theta(\varepsilon\indexj   - \mnui)
                          \right)
  \label{eq_betaspec_masses}
\end{eqnarray}
When this spectrum is convoluted with an experimental resolution
function which is much wider than the mass difference $|\mnui -
\mnuj|$ (which has always been the case so far),  the resulting
spectrum can be analyzed  in terms of a single mean squared
electron neutrino mass
\begin{equation}
  \mtwonue =  \sum_{\mathrm i} |U_{\mathrm ei}|^2 \cdot \mtwonui
\end{equation}
and eq. (\ref{eq_betaspec_finalstates}) applies again.

\begin{figure}
\centerline{\includegraphics[width=0.9\textwidth]{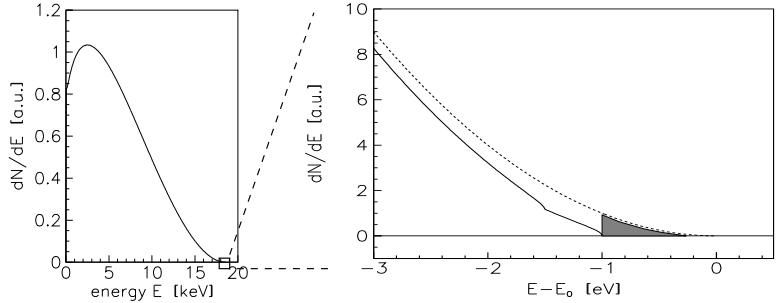}}
\caption{Tritium \bspec\ according to equation (\ref{eq_betaspec_masses})
  not taking into account the electronic final state distribution.
  Left: full \bspec , right: expanded region around tritium endpoint \ezero\
   for $\mnue = 0$ (dashed line) and for two neutrino mass
  states with arbitrarily chosen values of
  $\mnuone = 1.0$~eV, $|U_{\mathrm e1}|^2 = 0.7$,
  and $\mnutwo = 1.5$~eV, $|U_{\mathrm e2}|^2 = 0.3$,
  respectively (solid line).
  The gray shaded area corresponds to a fraction of $2 \cdot 10^{-13}$ of all
  tritium \bdec s.}
\label{fig_betaspec}
\end{figure}

The square root term of equation (\ref{eq_betaspec}) shows that
the neutrino mass influences the \bspec\ only at the upper end
below \ezero\ and its relative influence is vanishing  as function of
$\mtwonue / \varepsilon^2$ (see figure \ref{fig_betaspec}) leading far
below the endpoint to a small constant offset
proportional to $-\mtwonue$.

Figure \ref{fig_betaspec} defines the requirements of a direct
neutrino mass experiments which investigates a \bspec\ : The task
is to resolve the tiny change of the spectral shape due to the
neutrino mass in the region just below the endpoint \ezero\ ,
where the count rate is going to vanish. Therefore, high energy
resolution is required combined with large source strength and
acceptance as well as low background rate.

\section{Tritium $\beta$ Decay Experiments}
\label{sec_lablim_kin_nue_tri}

\begin{figure}
\centerline{\includegraphics[width=0.7\textwidth]{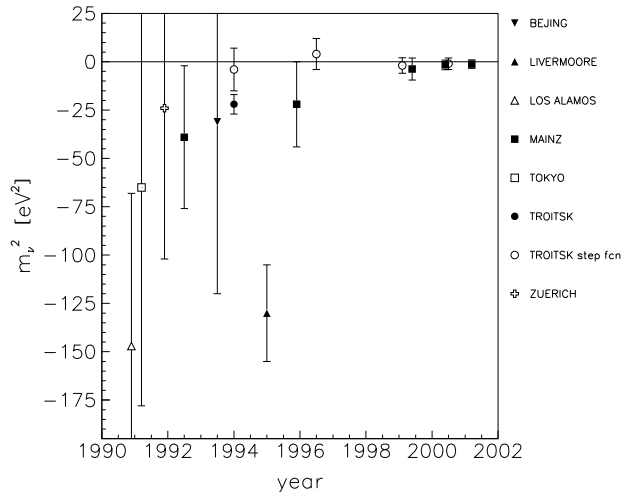}}
  \caption{Results of tritium \bdec\ experiments on the observable
  \mtwonue\
   over the last decade. The already finished
   experiments at Los Alamos, Z\"urich,
   Tokyo, Beijing and Livermore \protect\cite{LANLb,Zuerichb,Tokyo,Bejing,LLNL}
   used magnetic spectrometers, the
   experiments at Mainz and Troitsk 
    \protect\cite{weinheimer93,bonn97,weinheimer99,bonn01,belesev95,lobashev99,lobashev00} are using
   electrostatic spectrometers of the MAC-E-Filter type (see text).}
  \label{fig_tritiumexp}
\end{figure}

\subsection{Overview}
The majority of the published direct laboratory results on \mnue\
originates from
  the investigation of tritium \bdec , while one single result from $^{187}$Re has been
 reported at conferences \footnote{There are also results
  from investigations of electron capture \cite{elec_capture}
  and bound state \bdec\
  \cite{boundstate_bdec}, which are about 2 orders of magnitude less
  stringent on the neutrino mass.}.
  In the early nineties
  tritium \bdec\ experiments yielded controversially discussed results:
  Figure \ref{fig_tritiumexp} shows the final results of the
  experiments at Los Alamos National Laboratory 
  and Z\"urich together with the results from
  other more recent measurements with magnetic spectrometers
  at University of Tokio, Lawrence Livermore National
  Laboratory and at Bejing. The sensitivity
  on the neutrino mass have improved a lot compared to previous
  experiments but the values for the observable  \mtwonue\
  populated the unphysical negative \mtwonue\ region.
  In the case of two experiments significantly negative mass results were obtained.
  In 1991 and 1994 two new
  experiments started data taking at Mainz and at Troitsk,  which used
  a new type of electrostatic spectrometer, so-called MAC-E-Filters,
  which were again superior in energy resolution and luminosity with respect
  to the previous
  magnetic spectrometers. However, even their early data were confirming the
  large negative \mtwonue\ values of the Los Alamos and Livermore experiments
  when being analyzed over the last 500~eV of the \bspec\
  below the endpoint \ezero . Also a new feature was observed.
  The large negative values of \mtwonue\ disappeared when analyzing only
  small intervals below the endpoint \ezero\ (see also figure
  \ref{fig_mainzfits}). This effect, which could only be investigated by the high-resolution MAC-E filters,
   pointed towards
  an underestimated or missing energy loss process, seemingly to be present
  in all experiments. The only common feature of the various experiment
  seemed to be the calculations of the excitation energies
  $V\indexi$  of the daughter ions and their probabilities $W\indexi$.
  Different theory groups checked these calculations in detail.
  The expansion was calculated to one order further and new interesting
  insight into this problem was obtained, but no significant changes
  were found (see \cite{saenz} and references therein).

  Then the Mainz group found the origin of the missing
  energy loss process for its experiment. The Mainz experiment
  uses as tritium source a  film of molecular tritium
  quench-condensed onto aluminum or graphite substrates. Although the
  film was prepared as a homogenous thin film with flat surface, detailed studies
  showed that the film undergoes a temperature activated roughening transition
  into an inhomogeneous film by formation of microcrystals leading to
  unexpected large inelastic scattering probabilities.

  The Troitsk experiment on the other hand
  uses a windowless gaseous molecular tritium source, similar to the Los Alamos
  apparatus. Here, the influence of large angle scattering of electrons
  magnetically trapped in the tritium source was not considered in the
  first analysis. The Troitsk group stated, that if this effect is taken into account
  it gives a correction large enough to make the
  negative values for \mtwonue\ disappear.

  It is very likely that also for the experiments at Los Alamos and Livermore
  other experimental effects caused the negative values for \mtwonue .
  Any
  missed or underestimated experimental correction leads to a negative
  value for \mtwonue . This can be understood by the following consideration:
  for $\varepsilon \gg \mnue $,  eq. (\ref{eq_betaspec}) can be expanded into
  \begin{equation}
    \frac{dN}{dE} \propto \varepsilon^2 - \mtwonue/2
    \label{eq_betaspec_expanded}
  \end{equation}
  On the other hand the convolution of a \bspec\ (\ref{eq_betaspec}) with
  a Gaussian of width $\sigma$ leads to
  \begin{equation}
    \frac{dN}{dE} \propto \varepsilon^2 + \sigma^2
    \label{eq_betaspec_sigma}
  \end{equation}
  Therefore, in the presence of
  a missed experimental broadening with Gaussian width $\sigma$ one expects a shift
  of the result on \mtwonue\ of
  \begin{equation}
    \Delta \mtwonue \approx - 2 \cdot \sigma^2
    \label{eq_sigma_mtwonue}
  \end{equation}
  which gives rise to a negative value of \mtwonue .

  \subsection{MAC-E-Filter}
  The significant  improvement in the $\nu$--mass sensitivity by the Troitsk and the Mainz  experiments
  are due to MAC-E-Filters. This new type of spectrometer
  is based on early work by Kruit \cite{kruit} and was later
  redeveloped for the application to the tritium \bdec\
  at Mainz and Troitsk  independently \cite{pic92a,lob85}.
  The MAC-E-Filter combines high luminosity at low
  background and a high energy resolution. Both  features
 are essential to measure the
  neutrino mass from the endpoint region of a \bdec\ spectrum. The acronym MAC-E-Filter stands for
  \underline{M}agnetic \underline{A}diabatic \underline{C}ollimation followed by an
  \underline{E}lectrostatic \underline{Filter}.

  \begin{figure}
  \centerline{\includegraphics[angle=0,width=0.65\textwidth]{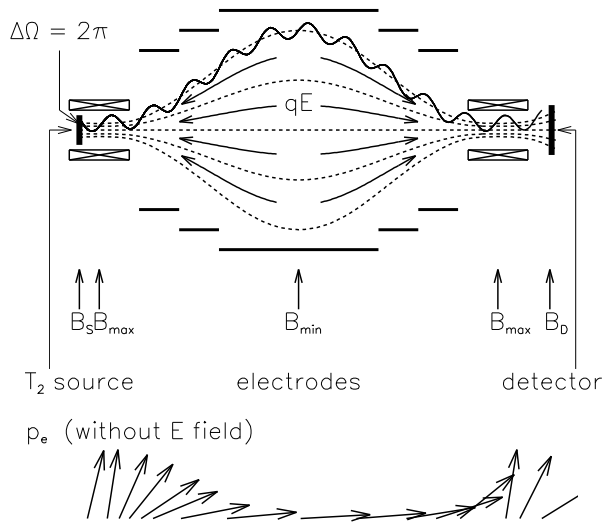}}
  \caption{Principle of the MAC-E-Filter. Top: experimental setup,
   bottom: momentum transformation due
   to adiabatic invariance of the magnetic orbit momentum $\mu$ in the
    inhomogeneous magnetic field.}
  \label{fig_mace}
  \end{figure}

  The main features of the MAC-E-Filter are illustrated in figure
  \ref{fig_mace}:
  two superconducting solenoids are producing a magnetic guiding field.
  The \belec s, which are starting from the tritium source
  in the left solenoid into the forward hemisphere, are
  guided magnetically on a cyclotron motion along the magnetic field lines
  into the spectrometer, thus resulting in an accepted solid
  angle of nearly $2 \pi$.
  On their way into the center of the spectrometer the magnetic
  field $B$ drops by several orders of magnitude. Therefore,
  the magnetic gradient force transforms most of the cyclotron energy $E_\perp$
  into longitudinal motion. This is illustrated in figure \ref{fig_mace} at the bottom
  by a momentum vector. Due to the slowly varying magnetic field the
  momentum transforms adiabatically keeping  the magnetic moment $\mu$
  constant
  (equation given in non-relativistic approximation):
  \begin{equation}
    \mu = \frac{E_\perp}{B} = const.
  \end{equation}
  This transformation can be summarized as follows:
  the \belec s, isotropically emitted at the source, are
  transformed into a broad beam of electrons flying almost parallel to the
  magnetic field lines.

  This parallel beam of electrons is  energetically analyzed by
  applying an
  electrostatic potential made up by a system of cylindrical electrodes.
  All  electrons, which have enough energy to pass the electrostatic barrier
  are reaccelerated and collimated onto a detector, all other electrons are reflected.
  Therefore the spectrometer acts as an integrating high-energy pass filter.
  The relative sharpness of this filter is only given by the ratio of the minimum
  magnetic field $B_{\rm min}$ in the analyzing plane in the middle
  and the maximum magnetic field between
  \belec\ source and spectrometer $B_{\rm max}$:
  \begin{equation}
    \frac{\Delta E}{E} = \frac{B_{\rm min}}{B_{\rm max}}
  \end{equation}
By scanning the electrostatic retarding potential the \bspec\ can be measured.

Both experiments at Mainz and Troitsk are using similar MAC-E-Filters,
which differ slightly in size: The diameter and length of the Mainz(Troitsk)
spectrometer are 1~m (1.5~m) and 4~m (7~m). The major differences between the
two setups are the tritium sources.

\subsection{The Troitsk Neutrino Mass Experiment}
The windowless gaseous tritium source of the Troitsk experiment
\cite{lobashev99}  is essentially a tube of 5~cm diameter filled
with \ttwo\ resulting in a column density of $\rho d \approx
10^{17}$~mole\-cu\-les$\rm /cm^2$. The source is connected to the
ultrahigh vacuum of the spectrometer by a series a differential
pumping stations.

From their first measurement in 1994 on the Troitsk group reported
the observation of a small, but significant
 anomaly in their experimental spectrum starting a few eV below the
$\beta$ endpoint \ezero . This anomaly appears as a sharp step of
the count rate \cite{belesev95}. Since a MAC-E-Filter is
integrating, this step should correspond to a narrow line in the
primary spectrum with a relative intensity of about $10^{-10}$ of
the total decay rate. In 1998 the Troitsk group reported that the
position of this line oscillates with a frequency of 0.5 years
between 5~eV and 15~eV below \ezero\ \cite{lobashev99}. In 2000
the anomaly did not follow the 0.5 year periodicity anymore but
still existed \cite{lobashev_erice01}. In total the Troitsk
experiment has taken 200 days of tritium data, in almost all of
the  runs this anomaly has been observed.

The reason for such an anomaly with these features is not clear.
Detailed investigations at Troitsk are continuing. In addition
synchronous measurements with the Mainz experiment were performed.
In 2001 the Troitsk group improved the differential pumping between
the gaseous tritium source and the spectrometer, lowered the
electric field strength in a critical region and improved the
vacuum. The first two runs of 2001  either gave no indication for
an  anomaly or only showed a small effect  with 2.5~mHz amplitude
if compared to the previous ones with amplitudes between 2.5~mHz
and 13~mHz. These findings also support the assumption that the
Troitsk anomaly is due to an still unknown experimental artefact
\cite{weinheimer02}.

Fitting a standard \bspec\ to the  data the Troitsk group obtained
significantly negative values of \mtwonue\ of -10 to -20 \evtwo\
(see filled circle in fig.  \ref{fig_tritiumexp}). Describing the
anomaly phenomenologically by adding a monoenergetic line, free in
amplitude and position, to a standard \bspec\ results in values of
\mtwonue\ compatible to zero \cite{lobashev99} (see also open
circles in fig. \ref{fig_tritiumexp}). After this correction the
average over all runs until 2001 amounts to
\cite{weinheimer02}:\\
\begin{equation}
  \mtwonue  = -2.3 \pm   2.5 \pm 2.0~\evtwo
\end{equation}
which corresponds - under the assumption that the run-by-run correction by
an additional line is correct - to an upper limit \cite{weinheimer02} of
\begin{equation}
  \mnue \leq 2.2 ~\ev ~~~~~{\rm (95~\%~C.L.)}
\end{equation}

\subsection{The Mainz Neutrino Mass Experiment}
  Mainz uses a
  film of molecular tritium quench-condensed onto a graphite substrate (HOPG).
  The film has a diameter of 17 mm and a typical thickness of 40 nm, which
  is measured by laser ellipsometry.
  The problem of the roughening transitions mentioned above have been investigated by the Mainz group
  in cooperation with the condensed matter
  group of P.~Leiderer at
  Konstanz/Germany using conversion electron spectroscopy and scattered light
  techniques on different hydrogen isotopes
  \cite{fleischmann1,fleischmann2}.
  The following  results were obtained:
  The roughening transition follows an Arrhenius-type law. Thus it
  cannot be avoided but its speed
  can drastically be slowed down by using lower temperatures.
  A \ttwo\ film at 2~K has a time
  constant of order 10~y \cite{fleischmann2}, i.e. much longer than a typical
  duration of a measurement.

\begin{figure}[t]
    \centerline{\includegraphics[angle=0,width=0.9\textwidth]{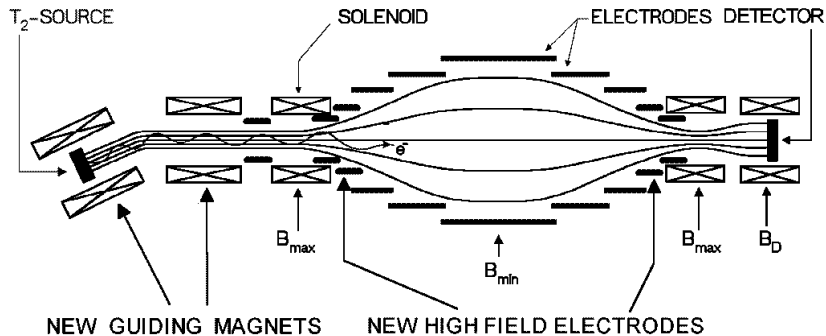}}
  \caption{The upgraded Mainz setup shown schematically, not to scale. 
    The outer diameter amounts to 1~m, the distance
    from source to detector is 6~m.}
  \label{fig_newsetup}
  \end{figure}

  In the years 1995-1997 the Mainz setup was  upgraded by a
  new cryostat providing temperatures of the tritium film below 2 K
  to avoid the roughening transition. Also  a new tilted pair of
  superconducting solenoids was installed (see figure
  \ref{fig_newsetup}).
  Consequently $\beta$ particles from the source
  are still guided magnetically into the spectrometer,
  whereas tritium molecules evaporating from the source
  are trapped on the bend of the LHe cold tube covered with graphite.
  This measure eliminated source correlated background and allowed to
  increase the source strength significantly.
  The upgrade of the Mainz setup was completed by the application
  of HF pulses on one of the electrodes in between measurements every
  20~s, and a full
  automation of the apparatus and remote control. The former improvement
  lowers and stabilizes the background, the latter one allows long--term
  measurements.

  \begin{figure}[t]
   \centerline{\includegraphics[angle=0,width=0.65\textwidth]{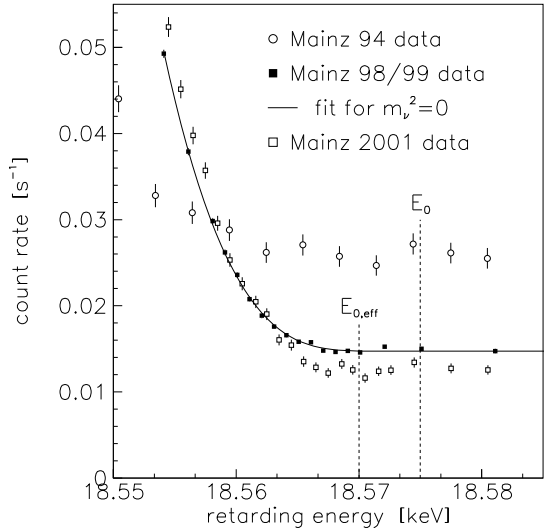}}
  \caption{Averaged count rate of the 1998/1999 data \cite{bonn01}
  (filled squares) with fit (line) 
  and of the 2001 data Q6--Q8 (open squares)\cite{kraus02} 
  im comparison with previous Mainz data
  from 1994 (open circles) \cite{bonn97}
  as function of the retarding
    energy near the endpoint \ezero , and effective endpoint $E_{0,eff}$.
    The position of
    the latter takes into account the width of the response function of the
    setup and
    the mean rotation-vibration excitation energy of the electronic
    ground state of the $\rm ^3HeT^+$ daughter molecule.}
  \label{fig_mainzdata}
  \end{figure}

  Figure \ref{fig_mainzdata}
  shows the endpoint region of the Mainz 1998 and 1999 data \cite{bonn01} in
  comparison with the former
  Mainz 1994 data \cite{bonn97}.
An improvement of the signal-to-background ratio
   by a factor 10
  as well as a significant enhancement of the statistical quality of the data
  is clearly visible. A fit with \mtwonue\  fixed to zero perfectly fits
  the latter data set over these last 15~eV of the \bspec .
This limits any persistent spectral anomaly in this range to an amplitude
below $10^{-3}$/s (as against a total flux of $10^8$/s entering the
spectrometer). A spectral anomaly, like the fluctuating anomaly reported
by the Troitsk group \cite{belesev95,lobashev99}, on the other hand,
reaches amplitudes up to $10^{-2}$/s.

  The main systematic uncertainties of the Mainz experiment
  are originating from the physics and the properties of the quench-condensed
  tritium film: the inelastic scattering of \belec s within the
  tritium film,
  the excitation of neighbor molecules due to the \bdec ,
  and the self-charging of the
  tritium film by  radioactivity.

  These systematic uncertainties were studied in detail by various investigations \cite{aseev,Erice,bornschein} and
  the knowledge of the corresponding corrections could be significantly improved.

  \begin{figure}[t]
   \centerline{\includegraphics[angle=0,width=0.65\textwidth]{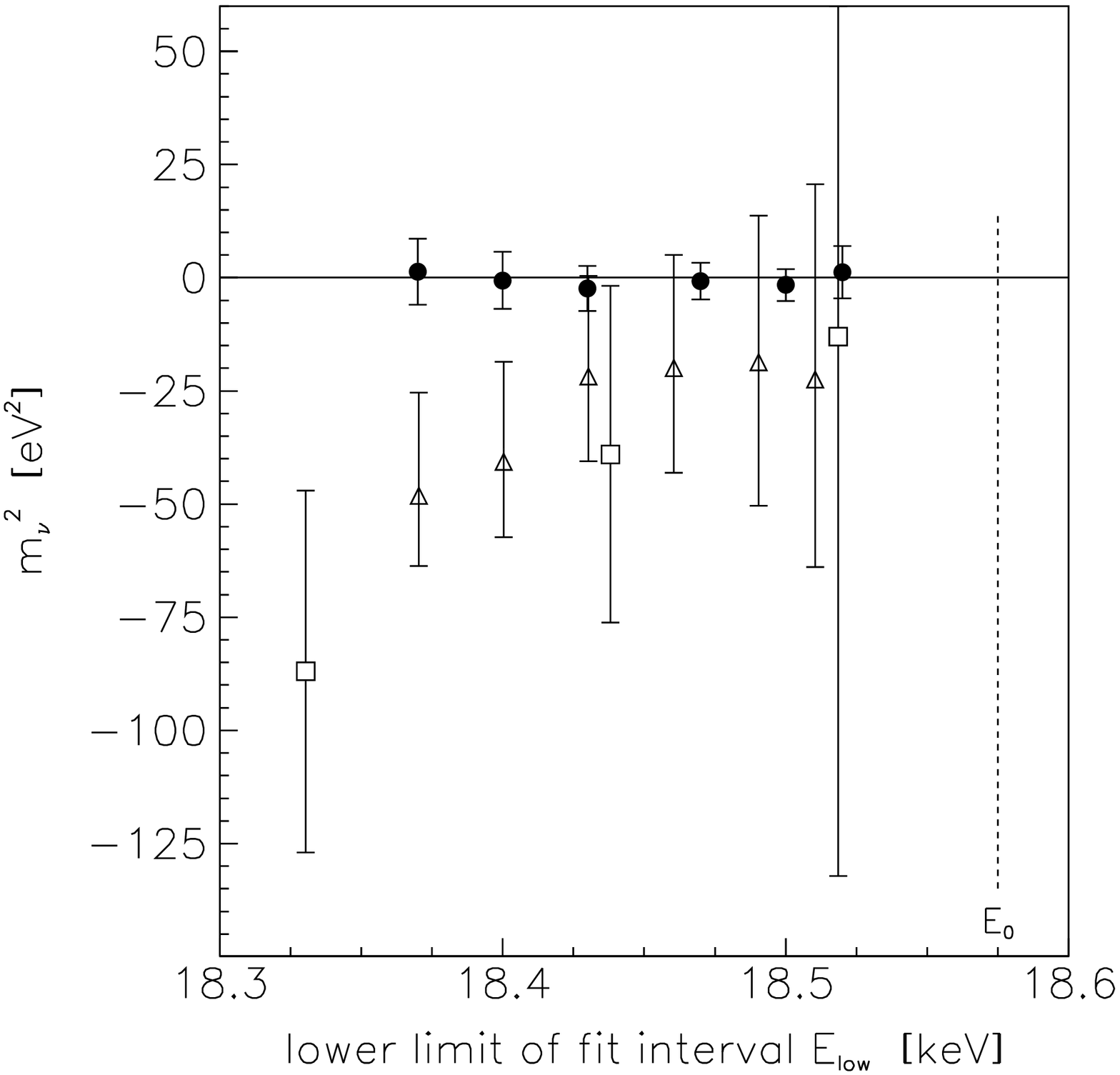}}
  \caption{Mainz fit results on \mtwonue\ in dependence
  on the lower limit of the fit interval (upper limit: always
  18.66~keV, well above \ezero ) for data from 1991
  \cite{weinheimer93} (open rectangles),
  from 1994  \cite{bonn97} (open triangles)
  and from the last four runs of 1998 and 1999 \cite{bonn01} (filled circles).}
  \label{fig_mainzfits}
  \end{figure}

  Figure \ref{fig_mainzfits} shows the fit results of the
  combined Mainz 1998 and 1999 data on \mtwonue\ as function of the
  lower limit of the fit interval.
  The monotonous trend towards negative values of \mtwonue\ for larger
  fit intervals, as  observed for the Mainz 1991 and 1994 data
  \cite{weinheimer93,bonn97}, has vanished. This shows
  that the dewetting of the \ttwo\ film from the graphite substrate
  \cite{fleischmann1,fleischmann2} indeed was the reason for
  this behavior. Now this effect is safely suppressed at the much
  lower temperature of the \ttwo\ film.
  The data do not show any indication for other residual distortions.
  The energy interval below the endpoint, yielding the
  smallest combined statistical and systematical
  uncertainty on the neutrino mass
  is obtained, corresponds  to the last 70 eV below the endpoint \ezero\
  and gives \cite{bonn01}
  \begin{equation}
  \label{eq_mainz_results9899}
  \mtwonue   =  -1.6 \pm 2.5 \pm 2.1~\evtwo
  \end{equation}
  which is compatible with a zero neutrino mass.
  Considering its uncertainties,
  this value corresponds to an upper limit on the electron neutrino mass of
  \cite{bonn01}:
  \begin{equation}
    \mnue \leq 2.2 ~\ev ~~~~~{\rm (95~\%~C.L.)}
  \end{equation}
Further data have been taken at Mainz in 2000, but suffering from
background problems
At the end of 2001 the Mainz group started another 3 month measurement
campaign. A very careful
maintenance and preparation of the whole setup was done. Especially all
parts which needs refreshment from time to time were replaced.
These were components related to the tritium source, the vacuum
and the high voltage system ({\it e.g.} the graphite substrate,
out-baking of all vacuum systems and re-activation of the
non-evaporable getter pumps, the oil of the high voltage divider,
and others). To check the various scanning hysteresis effects,
different scanning methods were tested, including a 50 times
slower scanning procedure. The summary of all these investigations
is: the Mainz experiment has achieved its most stable operation
ever in these two runs of 2001 -- no scanning direction
effect was observed anymore. The background
rate was about 13~mHz, lower than before and much more stable without
the need of repeating high voltage conditioning during a run (see fig.
\ref{fig_mainzdata}).

\begin{figure}[hbt]
   \centerline{\includegraphics[angle=0,width=0.65\textwidth]{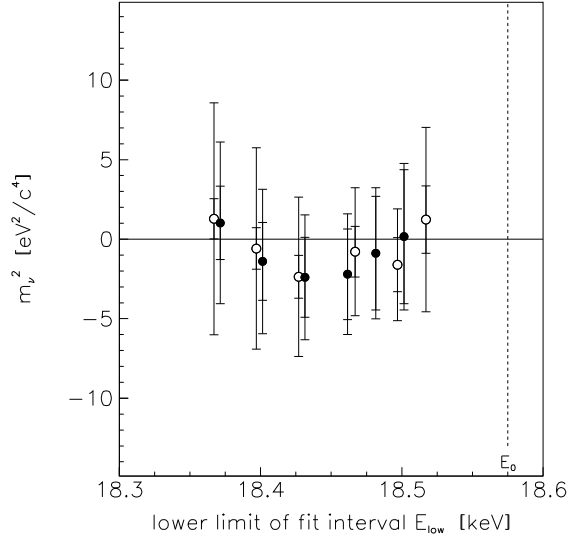}}
  \caption{Mainz fit results on \mtwonue\ as a function of the
  the lower boundary of the fit interval (the upper bound is fixed
  at 18.66~keV, well above \ezero ) for data from 1998 and 1999
  \protect\cite{bonn01} (open circles) and from the last runs of 2001
  (filled circles) \protect\cite{kraus02}.
  The error bars show the statistical uncertainties
  (inner bar) and the total uncertainty (outer bar).
  The correlation of data points for large fit intervals
  is due to the uncertainties of the systematic corrections, which are
  dominant for fit intervals with a lower boundary E$_{low}$\,$<$~18.5~keV.}
  \label{fig_mainzfits01}
\end{figure}

The neutrino squared masses obtained from the fit are very stable
and compatible with zero within their uncertainties and the
previous Mainz results (see figure \ref{fig_mainzfits01}). No indication
of a Troitsk-like anomaly or any residual problem in the Mainz
data were found. The result for the 2001 data of the last 70~eV of
the \bspec\ below the endpoint (E$_{\rm low}$=18.5~keV, see fig. 
\ref{fig_mainzfits01}) on \mtwonue\ is \cite{kraus02}:
\begin{equation}
\mtwonue = +0.1 \pm 4.2 \pm 2.0~ \evtwo 
\end{equation}
Combining this value with the one obtained from the data sets
Q5--Q8 from 1998 and 1999 over the last 70~eV (\ref{eq_mainz_results9899}) 
\cite{bonn01} gives
\begin{equation}
\mtwonue = -1.2 \pm 2.2 \pm 2.1~ \evtwo 
\end{equation}
which corresponds again to an upper limit \cite{kraus02,weinheimer02} of
\begin{equation}
\mnue < 2.2~ \ev \quad {\rm (95~\%~C.L.)}
\end{equation}

The inclusion of the high-quality data from 2001 improves the
Mainz sensitivity only marginally, showing that the Mainz
experiment has reached its sensitivity limit.

In spring 2002 the Mainz group has installed a new
electrode system to check new ideas which were deloped for the KATRIN 
experiment to avoid background and to
remove trapped particles \cite{mueller}. First measurements 
showed that the new
ideas indeed are reducing the background rate by a factor 3.

Indications for a ``Troitsk-like'' anomaly at Mainz were observed
only once in summer 1998. This single coincidence did not appear
in previous and later runs \cite{weinheimer99}. Of special
interest are the Mainz 2000 runs, although they were done under
less favored conditions. In particular, part of the data have been
taken in parallel with Troitsk. Whereas Troitsk  again has
indications for the anomaly \cite{lobashev_erice01} at Mainz no
indication for a Troitsk-like anomaly was found
\cite{otten_erice01}. In summary, the Mainz data clearly state
that the anomaly observed at Troitsk is caused by some
experimental artefact \cite{weinheimer02}.

\section{Future direct neutrino mass searches}
\label{sec_future}
As briefly discussed in section 1 the compelling evidence for non-zero
neutrino masses from atmospheric and solar neutrino experiments provides 
squared neutrino mass differences but no absolute neutrino masses.
These findings clearly demand
for the determination of the absolute neutrino mass scale as one of the
most important next steps in neutrino physics since the absolute neutrino
mass has strong consequences for astrophysics and cosmology as well as for
nuclear and particle physics as discussed in section 1. 

There exist different ideas and approaches 
to determine the absolute neutrino mass with
sub-eV sensitivity, which  will be briefly discussed in the following:

\begin{itemize}
\item {\bf Time-of-flight of supernova neutrinos}\\
The spread of the arrival time of supernova neutrinos on earth depends on 
neutrino energy and mass, thus allowing to extract the neutrino mass
by measuring arrival time and energy. 
A supernova, exploding within our galaxy, would give hundreds to 
thousand of neutrino events in the current underground neutrino 
detectors. Although this would exceed the statistics of 
the only supernova SN1987a observed so far by two orders of magnitude, 
the systematic uncertainty
connected with the not precisely known neutrino emission time spectrum does
not allow a sub-eV sensitivity on the neutrino mass.
\item {\bf Large scale structure}\\
The observation of the structure in the universe at different scales and
the angular distribution of the fluctuations of the cosmic microwave
background radiation allows to set constraints on the hot dark
matter content of the early universe \cite{LSS}. 
The relic neutrino density connects
this hot dark matter density to the neutrino mass.
Although the expected sensitivity on the neutrino mass for the near 
future is in the sub-eV range, the results derived this way 
are never model independent.
Reversing the arguements, there are strong degeneracies between the different 
parameters and its therefore very helpful to use information from
laboratory neutrino mass experiments to determine the other astrophysics
parameters more precisely.
\item {\bf Neutrinoless double \bdec }\\
The neutrinoless double \bdec\ is sensitive to the so-called
  ``effective'' neutrino mass
  \begin{equation}
    \label{eq_mee}
    \mee = | \sum_{\mathrm i} U_{\mathrm ei}^2 \cdot \mnui |\quad ,
  \end{equation}
  which is a coherent sum over all mass eigenstates contributing to the 
  electron neutrino with fraction  $U_{\mathrm ei}$. 
  The determination of \mee\ from the measurement of the neutrinoless
  double \bdec\ rate is complementary to the direct determination
  of the mass of the electron neutrino since \mee\ and \mnue\ can
  differ by the following reasons:
  \begin{enumerate}
    \item Double \bdec\ requires the neutrino to be a Majorana particle.
    \item The values $U_{\mathrm ei}^2$ in  eq. (\ref{eq_mee}) can have
       complex phases, which could lead to a partial cancellation of the
       different terms of the sum. Especially that the recent solar neutrino
       data point to large mixing opens this possibility
       \cite{smirnov}.
    \item The uncertainty of the nuclear matrix elements
       of neutrinoless double \bdec\ still contributes to the 
       uncertainty of \mee\ by about a factor of 2 .
     \item Non Standard Model processes, others than the exchange
       of a Majorana neutrino, could enhance the observed neutrinoless
       double \bdec\ rate without changing \mee . 
  \end{enumerate}
The lowest limit of $\mee < 0.35$~eV  is coming from the Heidelberg-Moscow
experiment using an array of semiconductor detectors of enriched $^{76}$Ge
\cite{klapdor_nosignal}.
Very recently part of the collaboration interpreted the data as a 
signal for neutrinoless double \bdec\ \cite{klapdor_signal}, 
a claim which raised discussions within 
the community. Not only to check this, but also 
for the reasons given above double \bdec\ experiments with much enhanced  
sensitivity are clearly needed. 
The proposed double \bdec\ experiments of the next generation 
aim for a sensitivity
on \mee\ in the range of 0.1~eV and below \cite{elliott_vogel}.
\item {\bf Rhenium cryogenic bolometer experiments}\\
Due to the complicated electronic
structure of \rhenium\ and its \bdec\ (compare to section
\ref{sec_lablim_kin_nue}) the advantage of the 7 times lower
endpoint energy \ezero\ of \rhenium\ with respect to tritium can
only be exploited if the $\beta$ spectrometer measures  the entire
released energy, except that of the neutrino. This situation can
be realized by using a cryogenic bolometer as the $\beta$
spectrometer, which at the same time contains the \bemit\
\rhenium\ (see figure \ref{fig_cryodetector}).

One disadvantage connected to this method is the fact that one
measures always the entire \bspec . Even for the case of the very
low endpoint energy of \rhenium , the relative fraction of events
in the last eV below \ezero\ is of order $10^{-10}$ only (compare
to figure \ref{fig_betaspec}). Considering the long time constant
of signal of a cryogenic bolometer (typically several hundred
$\mu$s) only large arrays of cryogenic bolometers can deliver the
signal rate needed.

Up to now two groups are working on \rhenium\ \bdec\ experiments
at Milano \cite{mubeta} and Genoa \cite{genua}. Although cryogenic
bolometers with an energy resolution of 5~eV have been produced
with other absorbers, this has yet not been reached for rhenium.
The two groups are  using different ways to produce the crystals.
The MANU2 experiment at Genoa succeeded in preparing crystals from
metallic rhenium. The group has reported a limit on \mnue\ of
26~\ev\ \cite{genua_limit}. The Genoa group understands their
measured spectra well and has seen for the first time the
oscillation pattern of the $\beta$ environmental fine structure
\cite{BEFS}, which describes the interference between the coherent
scattered outgoing electron wave function with the crystal similar
to the X-ray environmental fine structure (XEFS) known for X-ray
absorption edges. The Genoa group expects in the near future a
sensitivity on \mnue\ of 10~\ev . A significant further
improvement could be obtained by improving the energy resolution
of the crystals by using new super-conducting transition
thermometers. The MiBeta experiment of Milano uses  AgReO$_4$
crystals with a typical energy resolution of 35~eV. There is no
result on \mnue\ yet, the expected sensitivity is similar to the
one for the Genoa experiment.

 \begin{figure}
  \centerline{\includegraphics[angle=0,width=0.35\textwidth]{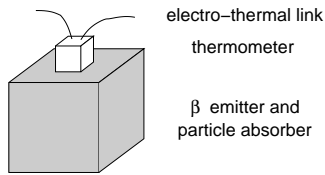}}
  \caption{Principle scheme of a cryogenic bolometer for direct
   neutrino mass measurements consisting of a $\beta$ emitting crystal, which
   serves at the same time the particle and energy absorber: The energy release
  $\Delta W$ gives rise  to a temperature increase
  $\Delta T = \Delta W/C$ via the heat capacity C, which is measured by a thermometer. The
  electric read-out wires of the thermometer link the whole bolometer
  to a thermal bath.}
  \label{fig_cryodetector}
 \end{figure}

To further improve the statistical
accuracy the operation of large arrays of
micro-calorimeters with better resolution is required. New techniques
are explored to enable these improvements.
The expected sensitivity on the neutrino mass
in the future is in the eV region \cite{genua_limit}.
\item {\bf Next generation tritium \bdec\ experiment}\\ 
Summarizing the discussion above clearly means that one or more 
next generation double \bdec\ experiments have to be performed due to
their very low sensitivity. But considering the complementariness 
of neutrinoless double \bdec\ and the direct neutrino mass determination
it is also
clear that a next generation direct mass search has to be done. None of the
alternative direct methods discussed above is able to provide 
a sub-eV sensitivity in the next decade. Therefore, it is straightforward
to explore which sensitivity could be achieved by
investigating the tritium \bdec\ spectrum near its endpoint 
with the very successful 
MAC-E-Filter as spectrometer.

Discussions between groups from Mainz, Karlsruhe and Troitsk led to the 
proposal for a next generation tritium \bdec\ experiment to be built at
Forschungszentrum Karlsruhe/Germany. This idea was strongly supported
by the community at the international workshop on ``Neutrino masses 
in the sub-eV range'' at Bad Liebenzell/Germany in early 2001
\cite{badliebenzell}. A strong collaboration including nearly the complete
worldwide expertise on tritium \bdec\ neutrino mass experiments has come
together and a 
letter of intent for the KATRIN experiment (KArlsruhe TRItium Neutrino
experiment) has been published \cite{KatrinLoI}. First funding for KATRIN
has been obtained. This experiment will be described in the next section 
in more detail.
\end{itemize}

\section{The KATRIN experiment}

\begin{figure}[tb]
\begin{center}
\includegraphics[width=0.9\textwidth]{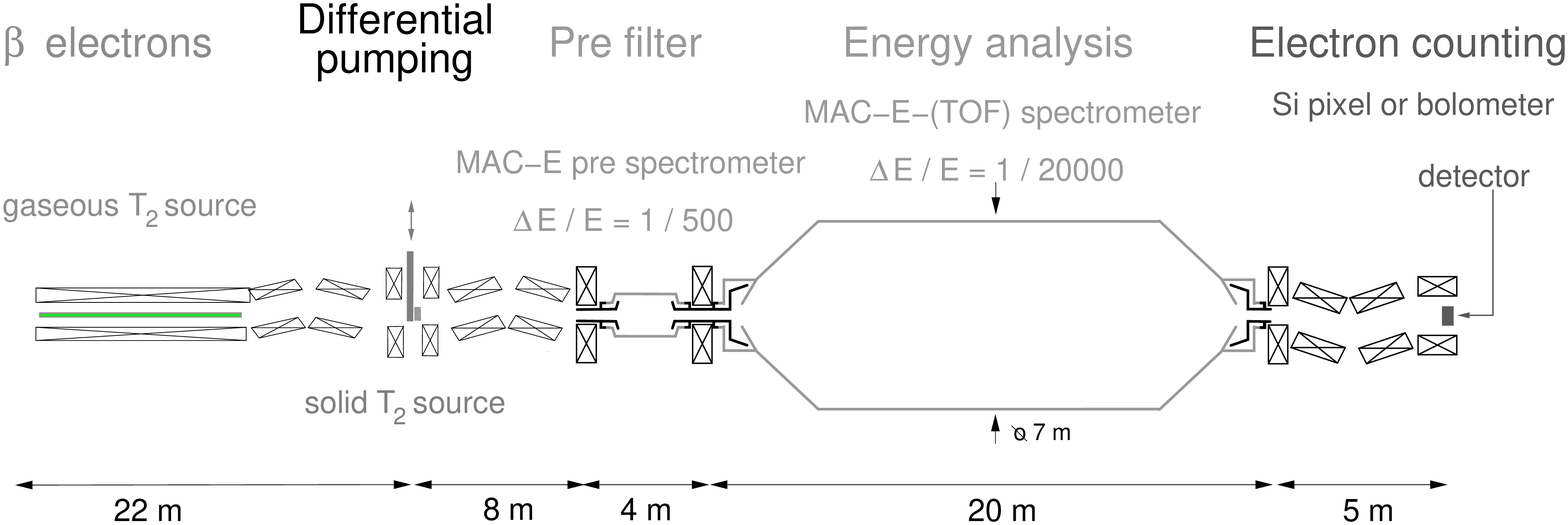}
\caption{Schematic view of the proposed next-generation tritium
  \bdec\ experiment KATRIN. The main components of the
  system comprise a windowless gaseous tritium source (WGTS), a
  alternative quench condensed tritium source (QCTS), a pre-spectrometer
  and a large electrostatic
  spectrometer with an energy resolution
  of 1\,eV. An electron transport system guides electrons from the
  \ttwo\ sources to the spectrometers, while eliminating all
  tritium molecules.
  \label{fig_katrin}}
\end{center}
\end{figure}

Figure 
 \ref{fig_katrin}) shows a schematic view of the proposed experimental 
 configuration.

The windowless gaseous tritium source (WGTS) allows the measurement of the
 endpoint region of the tritium \bdec\ and consequently  the determination
 of the neutrino mass with a minimum of systematic uncertainties 
from the tritium
 source. The WGTS will consist of a 10\,m long cylindrical tube of 70\,mm
diameter, filled with molecular tritium
gas of high isotopic purity ($> 95$~\%). The tritium gas will be
injected by a capillary at the middle of the tube and then diffuses over a
length of 5\,m to both end faces of the tube, resulting in a
source column density of about $(\rho d) \approx 5\cdot 10^{17}$ 
molecules/cm$^2$. This gives a near to  maximum count rate 
at the endpoint\footnote{One has to keep in mind, that increasing the source
column density further almost 
does not increase the count rate near the endpoint, 
since inelastic scattering of electrons on tritium molecules
has a threshold of about 12~eV \cite{aseev}. 
Any inelastically scattered electron
cannot contribute to the count rate in the interesting region of the
last 12~eV below the $\beta$ endpoint \ezero .}.
With these values the count rate is increased by factor 40 
with respect to the Troitsk experiment.

A  quench condensed tritium source (QCTS)
following the source concept of the Mainz experiment is considered as
a second alternative source, which has complementary systematics.

The electron transport system adiabatically guides \bdec\
electrons from the tritium sources to the spectrometer 
while at the same time eliminating any tritium flow towards
the spectrometer, which has to be
kept practically free of tritium for background reasons.
The first part of the transport system consists of a differential
pumping section with a tritium reduction of a factor $10^9$, the second part
of a liquid helium cold cryo-trapping section.

Between the tritium sources and the main spectrometer
a pre-spectrometer of MAC-E-Filter type will be inserted, acting
as an energy pre-filter to reject all \belec s except the
ones in the region of interest close to the endpoint \ezero .
This minimizes the chances of causing background by ionization of
residual gas in the main spectrometer.
As the designs of the
pre- and main spectrometer will be similar, the former
will act as a test facility for the larger main spectrometer \cite{flatt}. 
The design
and construction of the pre-spectrometer has already started.
 Especially important will be the following tests of\,: a) the technique
 to achieve an XUHV of below $10^{-11}$\,mbar, b) the concept of
 using the vacuum vessel itself as main electrode at high potential, 
and c) the new 
 electromagnetic design concepts to reduce background.

 A key component of the new experiment will be the large
 electrostatic main spectrometer with a diameter of 7\,m and an overall
 length of about 20\,m. This high resolution MAC-E-Filter will allow
 to scan the tritium \bdec\ endpoint with increased
 luminosity at a resolution of 1\,eV, which is a factor of 4
 better than present MAC-E-Filters at Mainz and Troitsk. The 100 times
larger analyzing plane  with respect to the Mainz experiment allows
the remaining factor 25 to be utilized to increase the source cross section 
and, correspondingly, the signal rate.

The detector requires high efficiency
for electrons at $\ezero = 18.6$~keV and low $\gamma$ background. A high energy
resolution of  $\Delta E<600$\,eV for 18.6\,keV electrons should suppress
background events at different energies.
The present concept of the detector is based on a large array of
about 1000 silicon drift detectors 
surrounded by low-level passive shielding and
an active veto counter to reduce background.
For a later stage of the experiment a segmented bolometer is
considered as a possible detector.

The difficulty in measuring a sub eV neutrino mass concerning the
statistics, namely the smallness of the interesting region
below the endpoint, turns into an advantage with respect to the 
systematic uncertainties due to energy thresholds for inelastic processes. 
First simulations with conservative assumptions
result in a total 1~$\sigma$ uncertainty of $\Delta \mtwonue = 0.08$\,eV$^2$ 
for a measurement
time of three years. If no finite neutrino mass is observed, 
the three year measurement leads  to an upper limit
of the mass  of 0.35\,eV at 90\% confidence. 
This sensitivity improves
the existing limits by almost one order of magnitude and also
demonstrates the discovery potential of KATRIN for an electron
neutrino mass in the sub-eV range.

\section{Summary}
Although neutrino oscillation experiments show that neutrinos have non-zero 
masses, the direct neutrino mass experiments have obtained only upper limits
yet.
The current tritium \bdec\ experiments at Mainz and Troitsk
are reaching their sensitivity limits. The Mainz upper 
limit on \mnue\ is 2.2~eV
at 95~\%~C.L. The Troitsk group gives the same limit under the 
assumption that an anomalous excess count rate
near the endpoint is described correctly. The
synchronous measurements at Mainz and Troitsk show that 
this ``Troitsk anomaly'' is an experimental artefact.

A mass determination with sub-eV sensitivity is clearly needed
to distinguish between hierarchical and degenerate neutrino mass models and
to clarify the role of neutrinos in the early universe.
The search for the neutrinoless double \bdec\ is one very important approach.
Complementary and equally important is a next
generation direct neutrino mass experiment. Discussing the different options
shows that this experiment has to be a large tritium \bdec\ experiment
using a MAC-E-Filter. Such an experiment is being prepared 
by the KATRIN collaboration.

\section*{Acknowledgments}
The author would like to thank the various 
collaborations for providing him kindly the presented informations and results.
Especially acknowledged are the 
valuable remarks by J.~Bonn, G.~Drexlin, Ch.~Kraus and E.~Otten. 
The work of the Mainz and the KATRIN experiments connected to the
author is supported by 
the German Bundesministerium f\"ur Bildung und Forschung under
contracts 06MZ866I/5 and 05CK2PD1/5.


\begin{thebibliography}{0}
%
%
%

%
%


\bibitem{pdg00} K. Hagiwara {\it et al.} (Particle Data Group), 
                Phys. Rev. \textbf{D66} (2002) 010001
\bibitem{pmu} K.~Assamagan {\it et al.}, Phys. Rev. D53 (1996) 6065
\bibitem{aleph} R.~Barate \etal\  Eur. Phys. J. {\bf C2} (1998) 3
\bibitem{robertson} R.G.H. Robertson, D.A. Knapp, Ann. Rev. Nucl. Sci. 38 
                    (1988) 185
\bibitem{holzschuh} E. Holzschuh, Rep. Prog. Phys. 55 (1992) 1035-1091
\bibitem{wilkerson}
J.F. Wilerson and R.G.H. Robertson in 
``Current Aspects Of Neutrino Physics,'', edited by D.~O.~Caldwell,
{\it  Berlin, Germany: Springer (2001)}
\bibitem{LANLb}     R.G.H. Robertson \etal , \PRL\ \textbf{67} (1991) 957
\bibitem{Zuerichb}  E. Holzschuh \etal , \PL\ \textbf{B287} (1992) 381
\bibitem{Tokyo}    H. Kawakami \etal , \PL\ \textbf{B256} (1991) 105
\bibitem{Bejing}    H.C. Sun \etal , CJNP {\bf 15} (1993) 261  
\bibitem{LLNL}     W. Stoeffl, D.J. Decman, \PRL\ \textbf{75} (1995) 3237
\bibitem{weinheimer93} Ch. Weinheimer {\it et al.}, Phys. Lett. 
  {\bf B300} (1993) 210
\bibitem{bonn97}  H. Backe \etal , Proc. of Neutrino 96, Helsinki/Finland, 
  June 1996, World Scien\-ti\-fic/Sin\-ga\-pure
\bibitem{weinheimer99} C. Weinheimer \etal , \PL\ \textbf{B460} (1999) 219
\bibitem{bonn01} J. Bonn \etal ,Nucl. Phys. B (Proc. Suppl.) {\bf 91} (2001), 
  273
\bibitem{belesev95} A.I. Belesev \etal , \PL\ {\bf B350} (1995) 263
\bibitem{lobashev99}   V.M. Lobashev \etal , \PL\ \textbf{B460} (1999) 227
\bibitem{lobashev00} V.M. Lobashev \etal , Nucl. Phys. B (Proc. Suppl.) 
      {\bf 91} (2000) 280
\bibitem{elec_capture} S.~Yasumi {\it et al.},
Phys.\ Lett.\ B {\bf 334} (1994) 229
\bibitem{boundstate_bdec}
M.~Jung {\it et al.},
Phys.\ Rev.\ Lett.\  {\bf 69} (1992) 2164.
\bibitem{saenz} A. Saenz \etal , Phys. Rev. Lett. {\bf 84} (2000) 242
\bibitem{kruit} P. Kruit and F.H. Read, J. Phys. E16 (1983) 313
\bibitem{pic92a}  A. Picard \etal , Nucl. Inst. Meth. {\bf B63} (1992) 345
\bibitem{lob85}  V.M. Lobashev, Nucl. Inst. and Meth. {\bf A240} (1985) 305
\bibitem{lobashev_erice01} V.M. Lobashev, Prog. Part. Nucl. Phys. 48 (2002) 
  123
\bibitem{weinheimer02} Ch. Weinheimer, talk at Int. Conf. on Neutrino Physics 
and Astrophysics, Munich/Germany, June 2002, {\it proc. in print}
\bibitem{fleischmann1} L. Fleischmann {\it et al.}, J. Low Temp. Phys. {\bf 119} (2000) 615
\bibitem{fleischmann2} L. Fleischmann \etal , Eur. Phys. J. {\bf B16} (2000) 521
\bibitem{aseev}    V.N. Aseev \etal , Eur. Phys. J. {\bf D10} (2000) 39
\bibitem{Erice} H. Barth \etal , Prog. Part. Nucl. Phys. {\bf 40} (1998) 353
\bibitem{bornschein} B. Bornschein, PhD thesis, Mainz University, 2000, {\it 
  submitted to J. Low. Temp. Phys.}
\bibitem{kraus02} Ch.~Kraus \etal , poster at Int. Conf. on Neutrino Physics 
and Astrophysics, Munich/Germany, June 2002, {\it proc. in print}
\bibitem{mueller} B. M\"uller, Th. Th\"ummler \etal , 
  poster at Int. Conf. on Neutrino Physics 
and Astrophysics, Munich/Germany, June 2002, {\it proc. in print}
\bibitem{otten_erice01} J. Bonn \etal , Prog. Part. Nucl. Phys. 48 (2002) 
  113
\bibitem{LSS} S. Hannestad,  talk at Int. Conf. on Neutrino Physics 
and Astrophysics, Munich/Germany, June 2002, {\it proc. in print}
\bibitem{smirnov} Y. Farzan, O.L.G. Peres, A. Yu. Smirnov, Nucl. Phys. B 
  {\bf 612} (2001) 59
\bibitem{mubeta} A. Nucciotti \etal , 
Nucl. Instr. Meth. A {\bf 444} (2000) 77 
\bibitem{genua} M.~Galeazzi \etal , 
 Phys.\ Rev.\ C {\bf 63} (2001) 014302
\bibitem{genua_limit} F. Gatti, Physics B (Proc. Suppl.) 91 (2001) 293
\bibitem{BEFS} F. Gatti \etal , Nature 397 (1999) 137
\bibitem{badliebenzell} Int. Workshop, 
Bad Liebenzell/Germany,\hfill \\ January 2001,\hfill \\ 
   {\small \tt http://www-ik1.fzk.de/tritium/liebenzell} 
\bibitem{klapdor_nosignal} 
H.~V.~Klapdor-Kleingrothaus {\it et al.},
Eur.\ Phys.\ J.\ A {\bf 12} (2001) 147.
\bibitem{klapdor_signal}
H.~V.~Klapdor-Kleingrothaus, A.~Dietz, H.~L.~Harney and I.~V.~Krivosheina
                  [Heidelberg-Moscow Collaborations],
Mod.\ Phys.\ Lett.\ A {\bf 16} (2001) 2409
[arXiv:hep-ph/0201231].
\bibitem{elliott_vogel} S.R. Elliott and P. Vogel, Annu. Rev. Nucl. Part. Sci. 52 (2002), hep-ph/0202264
\bibitem{KatrinLoI} A. Osipowicz \etal , hep-ex/0109033
\bibitem{flatt} B. Flatt and J. Wolf,  poster 
at Int. Conf. on Neutrino Physics 
and Astrophysics, Munich/Germany, June 2002, {\it proc. in print}
\end{thebibliography}
\end{document}